\documentclass[11pt,a4paper]{article}
\pdfoutput=1
\usepackage{graphicx}
\usepackage{jheppub}
\usepackage{latexsym,amsmath,amssymb,lmodern,float,url}
\usepackage{natbib}
\usepackage{color}
\usepackage{microtype}
\usepackage{slashed}
\usepackage{multirow}
\usepackage{bm}

\newcommand{\be}{\begin{equation}}
\newcommand{\ee}{\end{equation}}
\newcommand{\bea}{\begin{eqnarray}}
\newcommand{\eea}{\end{eqnarray}}

\newcommand{\ba}{\begin{array}}
\newcommand{\ea}{\end{array}}

\newcommand{\rjp}{R(J/\psi)}
\newcommand{\jp}{J/\psi}
\newcommand{\bc}{B_c^+}

\begin{document}
\title{Model-Independent Bounds on $R(J/\psi)$}
\author[a]{Thomas D. Cohen}
\author[a]{Henry Lamm}
\author[b]{and Richard F. Lebed}

\emailAdd{cohen@umd.edu}
\emailAdd{hlamm@umd.edu}
\emailAdd{Richard.Lebed@asu.edu}

\affiliation[a]{Department of Physics, University of Maryland, College
Park, MD 20742, USA}
\affiliation[b]{Department of Physics, Arizona State University, Tempe,
AZ 85287, USA}

\abstract{
We present a model-independent bound on $R(J/\psi) \! \equiv \!
\mathcal{BR} (B_c^+ \rightarrow J/\psi \, \tau^+\nu_\tau)/$
$\mathcal{BR} (B_c^+ \rightarrow J/\psi \, \mu^+\nu_\mu)$.  This bound
is constructed by constraining the form factors through a combination
of dispersive relations, heavy-quark relations at zero-recoil, and the
limited existing determinations from lattice QCD\@.  The resulting
95\% confidence-level bound, $0.20\leq R(J/\psi)\leq0.39$, agrees with
the recent LHCb result at $1.3 \, \sigma$, and rules out some
previously suggested model form factors.  }

\maketitle
\section{Introduction}

\begin{figure}[ht]
 \includegraphics[width=\linewidth]{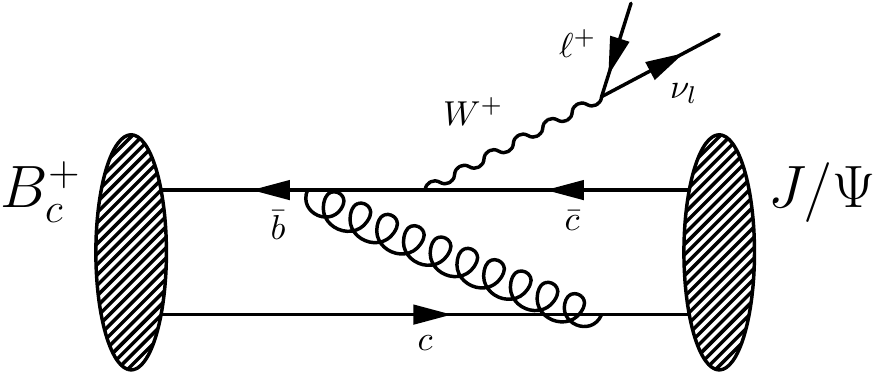}
 \caption{\label{fig:bcdecay}Schematic picture of the $\bc\rightarrow
  \jp \, \ell^+ \nu_\ell$ process.}
\end{figure}

Within the Standard Model, {\it lepton universality\/} is broken only
by the Higgs interaction, but the discovery of neutrino masses implies
that at least one, and potentially several, relevant forms of
beyond-Standard Model modification exist.  The ratios of semileptonic
heavy-meson decay branching fractions to distinct lepton
flavors represent a group of observables particularly sensitive to new
physics, because the QCD dynamics of the heavy-meson decays
decouples from the electroweak interaction at leading
order:
\begin{equation}\label{eq:matel}
|\mathcal{M}_{\bar{b}\rightarrow\bar{c} \, \ell^+ \nu_\ell}|^2=
\frac{L_{\mu\nu}H^{\mu\nu}}{q^2-M_W^2}+\mathcal{O}(\alpha,G_F)\,.
\end{equation}
This expression implies that the ratios of semileptonic heavy-meson
decay branching fractions can differ from unity at this
level of precision only due to kinematic factors.  Measurements from
BaBar, Belle, and LHCb of the ratios $R(D^{(*)})$ for
heavy-light meson decays $B \! \rightarrow \!  D^{(*)}\ell\bar{\nu}$
with $\ell \! = \! \tau$ to those with $\ell \! = \! \mu$
or $e$ (or their average) exhibit tension with theoretical
predictions.  The HFLAV averages~\cite{Amhis:2016xyh} of
the experimental results $R(D^*) \! = \!
0.306(13)(7)$~\cite{Lees:2012xj,Lees:2013uzd,Huschle:2015rga,
Sato:2016svk,Aaij:2015yra,Hirose:2016wfn,Aaij:2017uff,Aaij:2017deq,
Hirose:2017dxl} and $R(D) \! = \!
0.407(39)(24)$~\cite{Lees:2012xj,Lees:2013uzd,Huschle:2015rga}
represent a combined 3.8$\sigma$
discrepancy~\cite{Amhis:2016xyh} from the HFLAV-suggested
Standard-Model value of $R(D^*) \! = \! 0.258(5)$~\cite{Amhis:2016xyh}
obtained by an
averaging~\cite{Bernlochner:2017jka,Bigi:2017jbd,Jaiswal:2017rve} that
utilizes experimental data, lattice QCD results, and
heavy-quark effective theory, and from $R(D) \! = \!
0.300(8)$~\cite{Aoki:2016frl}, which is an average of lattice QCD
results~\cite{Lattice:2015rga,Na:2015kha}, as well as a value $R(D) \!
= \! 0.299(3)$ obtained by also including experimental data supplemented by heavy-quark effective theory~\cite{Bigi:2016mdz}.
In light of this tension, the LHCb Collaboration has measured the
rates for the heavy-heavy semileptonic meson decays $\bc \!
\rightarrow \! \jp \, \ell^+\nu_\ell$ (Fig.~\ref{fig:bcdecay}) in the
$\ell \! = \! \tau,\mu$ channels, finding $R(\jp) =
0.71(17)(18)$~\cite{Aaij:2017tyk}.

At present, only model-dependent calculations of $\rjp$ exist
(collected in
Table~\ref{tab:models})~\cite{Anisimov:1998uk,Kiselev:1999sc,
Ivanov:2000aj,Kiselev:2002vz,Ivanov:2006ni,Hernandez:2006gt,
Qiao:2012vt,Wen-Fei:2013uea,Rui:2016opu,Dutta:2017xmj,
Liptaj:2017nks,Watanabe:2017mip,Issadykov:2018myx,Tran:2018kuv}.
Although most models' central values cluster in LHCb's quoted theory
range of $0.25$--$0.28$, one notes a wide spread in their estimated
uncertainty.  We take as a reasonable estimate of the model range $0
\! < \! \rjp \! < \! 0.48$, the union of the 95\% confidence
levels (CL) of the reported theoretical uncertainties, which in turn
typically account only for parameter fitting.  These results rely upon
approximations such as nonrelativistic reduction, constituent quarks,
or perturbative QCD to obtain transition form factors between the
heavy-heavy $B_c^+$ and $\jp$ mesons.  Without a clear understanding
of the systematic uncertainties these assumptions introduce, the
reliability of these predictions is suspect.

In this paper, we present the first model-independent constraint, a
95\% CL bound of $0.20\leq\rjp\leq0.39$ within the Standard Model, in
which uncertainties are all quantifiable.  In order to obtain this
result, we begin in Sec.~\ref{sec:sm} with a discussion of the $V \! -
\! A$ structure of the Standard Model and the form factors.  In
Sec.~\ref{sec:hqss} we explain how heavy-quark spin symmetry can be
applied at the zero-recoil point to relate the form factors, using the
method of~\cite{Kiselev:1999sc}.  The initial lattice-QCD results of
the HPQCD Collaboration~\cite{Colquhoun:2016osw,ALE} for two of the
transition form factors are discussed in Sec.~\ref{sec:lat}\@.  The
dispersive analysis framework utilized to constrain the form factors
as functions of momentum transfer is presented in Sec.~\ref{sec:da}\@.
The results of our analysis, as well as future projections for the
bound, appear in Sec.~\ref{sec:results}, and we conclude in
Sec.~\ref{sec:con}.

\section{Structure of $\langle \jp \, |(V \! - \! A)^\mu|\bc
\rangle$}\label{sec:sm}

In the Standard Model, the factorization of Eq.~(\ref{eq:matel}) into
a leptonic and a hadronic tensor reduces the problem of calculating
$\rjp$ to the computation of the hadronic matrix element $\langle
\jp \, |(V \! - \! A)^\mu|\bc\rangle$.  Using this factorization,
the hadronic matrix element can be written in terms of four transition
form factors.  These form factors enter the matrix element in
combination with the meson masses, $M \! \equiv \! M_{\bc}$ and $m \!
\equiv \!  M_{\jp}$, the corresponding meson momenta $P^\mu$ and
$p^\mu$, and the polarization $\epsilon^\mu$ of the $\jp$.  The form
factors themselves depend only upon $q^2 \! = \! (P-p)^2$, the squared
momentum transfer to the leptons.  A number of form-factor
decompositions exist in the literature; one common
set~\cite{Wirbel:1985ji} used in
lattice-QCD~\cite{Colquhoun:2016osw} and model calculations is
given by $V(q^2),A_i(q^2)$, $i \! = \! 0,1,2,3$:

\begin{align}\label{eq:hadme}
 \langle \jp(p,\epsilon)|(V-A)^\mu|\bc(P)\rangle=&
\frac{2i\epsilon^{\mu\nu\rho\sigma}}{M+m}
\epsilon^{*}_{\nu}p_{\rho}P_{\sigma}V(q^2)-(M+m)
\epsilon^{*\mu}A_1(q^2)\nonumber\\
 &+\frac{\epsilon^{*}\cdot q}{M+m}(P+p)^\mu A_2(q^2)
+2m\frac{\epsilon^{*}\cdot q}{q^2}q^{\mu}A_3(q^2)\nonumber\\
 &-2m\frac{\epsilon^{*}\cdot q}{q^2}q^{\mu}A_0(q^2) \, ,
\end{align}
where $q^\mu \! \equiv \! (P-p)^\mu$.  While we have exhibited five
form factors, only four are independent.  In the physical set,
$A_0(q^2)$ is defined as the unique form factor that couples to
timelike virtual $W$ polarizations ($\propto \! q^\mu$), while
$A_3(q^2)$ is simply a convenient shorthand for a combination
appearing in intermediate stages of calculations, and in fact
satisfies
\begin{equation}\label{eq:a3}
 A_3(q^2)=\frac{M+m}{2m}A_1(q^2)-\frac{M-m}{2m}A_2(q^2) \, .
\end{equation}
Furthermore, the finiteness of Eq.~(\ref{eq:hadme}) as $q^2 \! \to \!
0$ requires $A_3(0) \! = \! A_0(0)$, which proves useful in
constructing our bounds.  In what follows, we also use the notation
$t \! \equiv \! q^2$, and define two important kinematic points,
$t_\pm \! = \! (M\pm m)^2$.

Using Eq.~(\ref{eq:hadme}) or an equivalent basis, model predictions
include uncontrolled approximations for the form factors.  Some
models construct wave functions for the two mesons, while others
attempt to compute a perturbative distribution amplitude at $q^2 \!
\rightarrow \! 0$ and then extrapolate to larger values with some
functional form.  In addition, some models do not respect form-factor
relations, such as the heavy-quark spin-symmetry relations discussed
below.  Due to these issues, the good agreement seen between the
model predictions may more reflect the theoretical prejudice in
modeling than a genuine estimate of the true Standard-Model value.
\begin{table}
\caption{\label{tab:models}Model predictions of $\rjp$ classified by
method, which are abbreviated as: constituent quark model (CQM),
relativistic quark model (RCQM), QCD sum rules (QCDSR),
nonrelativistic quark model (NRQM), nonrelativistic QCD (NRQCD), and
perturbative QCD calculations (pQCD).}
\begin{center}
\begin{tabular}
{l| c c}
\hline\hline
Model & $R_{theory}$ & Year\\
\hline
CQM~\cite{Anisimov:1998uk} & 0.28  &1998   \\
QCDSR~\cite{Kiselev:1999sc}&
$0.25^{+0.09}_{-0.09}$&1999\\
RCQM~\cite{Ivanov:2000aj} &0.26   & 2000  \\
QCDSR~\cite{Kiselev:2002vz} &0.25   &2003   \\
RCQM~\cite{Ivanov:2006ni} & 0.24  &2006  \\
NRQM~\cite{Hernandez:2006gt} &$0.27^{+0.02}_{-0}$   &2006   \\
NRQCD~\cite{Qiao:2012vt} & $0.07^{+0.06}_{-0.04}$ &2013   \\
pQCD~\cite{Wen-Fei:2013uea} & $0.29^{+0.09}_{-0.09}$  &2013   \\
pQCD~\cite{Rui:2016opu} & $0.30^{+0.11}_{-0.08}$ & 2016\\
pQCD~\cite{Dutta:2017xmj} &$0.29^{+0.07}_{-0.07}$ &2017\\
CQM~\cite{Liptaj:2017nks} &0.24   &2017   \\
pQCD~\cite{Watanabe:2017mip} &0.283$_{-0.048}^{+0.048}$ & 2017\\
CQM~\cite{Issadykov:2018myx} &$0.24^{+0.07}_{-0.07}$   &2018   \\
RCQM~\cite{Tran:2018kuv}& $0.24$&2018\\
\hline
Range & 0--0.48 & --\\
\hline
\end{tabular}
\end{center}
\end{table}

While this decomposition is useful for lattice QCD, it is not the best
decomposition for the dispersive analysis.  The second convention we
use is the helicity basis, which exchanges the form factors $V,A_i$
for $g$, $f$, $\mathcal{F}_1$, and $\mathcal{F}_2$.\footnote{Strictly
speaking, $\mathcal{F}_{1,2}$ are helicity amplitudes (in conventional
notation~\cite{Richman:1995wm}, proportional to $H_{0,t}$,
respectively), while $f,g$ are two linear combinations of them: $H_\pm
(t) \! = \! f(t) \mp k \sqrt{t} g(t)$, where $k$ is defined in
Eq.~(\ref{eq:kdef}).}  They are related by
\begin{eqnarray}
 g&=&\frac{2}{M+m} V \, , \nonumber\\
 f&=&(M+m)A_1 \, , \nonumber\\
 \mathcal{F}_1&=&\frac{1}{m}\left[-\frac{2k^2 t}{M+m} A_2
-\frac{1}{2}(t-M^2+m^2)(M+m)A_1\right] \, , \nonumber\\
 \mathcal{F}_2&=&2A_0 \, , \label{eq:FFrelns}
\end{eqnarray}
where, in terms of the spatial momentum $\bm{p}$ of the $\jp$ in the
$\bc$ rest frame,
\begin{equation} \label{eq:kdef}
k \equiv M \sqrt{\frac{\bm{p}^2}{t}}
= \sqrt{\frac{(t_+-t)(t_--t)}{4t}} \, .
\end{equation}

Setting $t=t_-$, we see that $\mathcal{F}_1(t_-)=(M-m)f(t_-)$.  In
this decomposition, the constraint $A_3(0) \! = \! A_0(0)$ reads
$\mathcal{F}_1(0)= \frac 1 2 (M^2-m^2)\mathcal{F}_2(0)$.  The
differential cross section for the semileptonic decay then reads
\begin{align}
\label{eq:difcof}
 \frac{d\Gamma}{dt}=&\frac{G_F^2|V_{cb}|^2}{192\pi^3M^3}
\frac{k}{t^{5/2}}\left(t-m_\ell^2\right)^2
\times \left\{ \left(2t+m_\ell^2\right)\left[2t|f|^2+
|\mathcal{F}_1|^2+2k^2t^2|g|^2\right] 
+3m_\ell^2k^2t|\mathcal{F}_2|^2
\right\} \, .
\end{align}

Inspecting Eq.~(\ref{eq:difcof}), one can see that in the light
leptonic channels ($\ell=e,\mu$), the contribution to the
$\mathcal{F}_2(t)$ can be neglected, while in the $\tau$ channel it
cannot.  As seen below, the uncertainty in our bound on $\rjp$ is
dominated by the unknown form factor $\mathcal{F}_2(t)$.

\section{Heavy-Quark Spin Symmetry}\label{sec:hqss}

Decays of heavy-light $Q\bar q$ systems possess enhanced symmetries in
the heavy-quark limit because operators that distinguish between heavy
quarks of different spin and flavor are suppressed by $1/m_Q$, and
their matrix elements vanish when $m_Q\rightarrow \infty$.
Consequently, all transition form factors $\langle Q'\bar q \,
|\mathcal{O}|Q \bar q\rangle$ in this limit are proportional to a
single, universal Isgur-Wise function
$\xi(w)$~\cite{Isgur:1989vq,Isgur:1989ed}, whose momentum-transfer
argument is $w$, the dot product of the initial and final heavy-light
hadron 4-velocities, $v^\mu \equiv p_M^\mu / M$ and $v'^\mu\equiv
p_m^\mu / m$, respectively:
\begin{equation} \label{eq:wdef}
w \equiv v \cdot v' = \gamma_m = \frac{E_m}{m} =
\frac{M^2 + m^2 - t}{2Mm} \, .
\end{equation}
At the zero-recoil point $t \! = \! (M \! - \! m)^2$ or $w \! = \! 1$,
the daughter hadron $m$ is at rest with respect to the parent $M$.
Indeed, one notes that $w$ equals the Lorentz factor $\gamma_m$ of $m$
in the $M$ rest frame.  The maximum value of $w$ corresponds to the
minimum momentum transfer $t$ through the virtual $W$ to the lepton
pair, which occurs when the leptons are created with minimal energy,
$t \! = \!  m_\ell^2$.

In heavy-light systems, the heavy-quark approximation corresponds to a
light quark bound in a nearly static spin-independent color field.  In
the weak decay $Q\rightarrow Q'$ between two very heavy quark flavors,
as the momentum transfer to the light quark
$t\rightarrow0$, $q$ no longer changes states, and therefore the wave
function of this light spectator quark remains unaffected.  One thus
concludes that $\xi(1) \! = \! 1$ at the zero-recoil (Isgur-Wise)
point, yielding a absolute normalization for the form factors.  These
results are accurate up to corrections of $\mathcal{O}(\Lambda_{\rm
QCD}/m_{Q'})$.

In the decay $\bc \! \rightarrow \! \jp$, the spectator light quark is
replaced by another heavy quark, $c$.  This substitution results in a
pair of related effects on the enhanced symmetries of the heavy-quark
limit~\cite{Jenkins:1992nb}.  First, the difference between the
heavy-quark kinetic energy operators produces energies no longer
negligible compared to those of the spectator $c$, and this effect
spoils the flavor symmetry in heavy-heavy systems.  Furthermore, the
spectator $c$ receives a momentum transfer from the decay of $\bar{b}
\! \to \! \bar{c}$ of the same order as the momentum imparted to the
$\bar{c}$, so one cannot justify a normalization of the form factors
at the zero-recoil point based purely upon symmetry.

While the heavy-flavor symmetry is lost, the separate spin symmetries
of $\bar{b}$ and $\bar{c}$ quarks remain, with an additional spin
symmetry from the heavy spectator $c$.  Furthermore, the presence of
the heavy $c$ suggests a system that is closer to a nonrelativistic
limit than heavy-light systems.  In the $\bc \! \rightarrow \! \jp$
semileptonic decays, one further finds that
\begin{eqnarray}
  w_{\rm max} & = & w(t \! = \! m_\ell^2) =
 \frac{M^2 +m^2 - m_\ell^2}{2M m} 
\approx 1.28 \ (\mu) , \ 1.20 \ (\tau) \, , \nonumber \\
  w_{\rm min} & = & w \left(t \! = \! (M \! - \! m)^2 \right) = 1 \, ,
\label{eq:wlimits}
\end{eqnarray}
suggesting that an expansion about the zero-recoil point may still be
reasonable.  Together, the spin symmetries imply that the four form
factors are related to a single, universal function $h$ ($\Delta$ in
Ref.~\cite{Jenkins:1992nb}), but only at the zero-recoil point, and no
symmetry-based normalization for $h$ can be
derived~\cite{Jenkins:1992nb}.

Using the trace formalism of~\cite{Falk:1990yz}, in
Ref.~\cite{Jenkins:1992nb} it was shown how to compute the relative
normalization between the four $\bar{Q}q\rightarrow \bar{Q}'q$ form
factors near the zero-recoil point [{\it i.e.}, where the spatial
momentum transfer to the spectator $q$ is $\mathcal{O}(m_q)$].
Using these relations, $h$ was derived for a color-Coulomb potential
in Ref.~\cite{Jenkins:1992nb}.  This approximation was improved in
Ref.~\cite{Colangelo:1999zn}, where a constituent quark-model
calculation of $\mathcal{BR}(\bc \! \rightarrow \! \jp \, \ell^+
\nu_\ell)$ for $\ell=e,\mu$ but not $\tau$, was performed.  The
heavy-quark spin-symmetry relations  of
Ref.~\cite{Jenkins:1992nb} were generalized in~\cite{Kiselev:1999sc}
to account for a momentum transfer to the spectator quark occurring at
leading order in NRQCD,  specifically, to the case $v \!
\neq \! v^\prime$ but $w \! \to \! 1$.  We reproduce here the
relations of~\cite{Kiselev:1999sc}, where the form factors
$g(w=1),\mathcal{F}_1(w=1),\mathcal{F}_2(w=1)$ are related to $f(w=1)$
by
\begin{eqnarray}\label{eq:hqss}
 g(w=1)& = & \frac{2\rho+(1+\rho)\sigma}{4M^2r\rho}f(w=1) \, ,
\nonumber \\
 \mathcal{F}_1(w=1) & = & M(1-r)f(w=1) \, , \nonumber\\
 \mathcal{F}_2(w=1) & = & \frac{2(1+r) \rho + (1-r)(1-\rho)\sigma}
{4Mr\rho}f(w=1) , \ \
\end{eqnarray}
where $r \! \equiv \! m/M$, $\rho \! \equiv \! m_{Q'} \! /m_Q$, and
$\sigma \! \equiv \! m_q/m_Q$.  These relations reproduce the standard
Isgur-Wise result~\cite{Isgur:1989vq,Isgur:1989ed,Boyd:1997kz} when
$\sigma \! = \! 0$.  The relation between
$\mathcal{F}_1(w=1)$ and $f(w=1)$ follows directly from the definition
of Eq.~(\ref{eq:FFrelns}), independent of heavy-quark symmetries.
Terms that break these relations should be $\mathcal{O}(m_c/m_b, \,
\Lambda_{\rm QCD}/m_c)\approx30\%$, and we allow conservatively for up
to 50\% violations.  The heavy-quark spin symmetry further relates the
zero-recoil form factors of $\bc \! \rightarrow \! \jp$ to those of
$\bc \! \rightarrow \! \eta_c$, which will be useful in the future to
obtain further constraints.

In analogy with the heavy-light systems, we can enforce a further
constraint from heavy-quark symmetries.  The universal form factor $h$
represents the overlap element of the initial and final states, and
therefore should be maximized at $w \! = \! 1$.  This
statement is an assumption, but a very mild one: In the heavy-light
system, the slope of the Isgur-Wise function is rigorously negative at
$w \! = \! 1$~\cite{Bjorken:1990hs,Isgur:1990jf,Uraltsev:2000ce}, and
it would indeed be very surprising if the same did not hold for the
form factors of heavy-heavy mesons, which are even more similar to
idealized quark-model states.

\section{Lattice QCD Results}\label{sec:lat}

The state-of-the-art lattice QCD calculations for $\bc \! \rightarrow
\! \jp$ are limited to preliminary results from the HPQCD
Collaboration for $V(q^2)$ at two $q^2$ values and $A_1(q^2)$ at three
$q^2$ values~\cite{Colquhoun:2016osw,ALE}.  These results were
obtained using 2+1+1 HISQ ensembles, in which the smallest lattice
spacing is $a\approx0.09$~fm, and the $b$ quark is treated via NRQCD,
and are reproduced in Fig.~\ref{fig:latff}.  For
$q^2=t_-,0$ $A_1(q^2)$ has also been computed on coarser lattices and
for lighter dynamical $b$-quark ensembles, in order to
check the accuracy and assess the uncertainty of the $a\approx0.09$~fm
NRQCD results.  At present, there are no lattice results for $A_0(q^2)
\! = \! \frac 1 2 \mathcal{F}_2 (q^2)$ or $A_2(q^2)$.  Below, we show
that the most desirable piece of new information from the lattice is a
computation of $A_0(0)$, which could cut our uncertainties in half.

\section{Dispersive Relations}\label{sec:da}

In this work we derive constraints on the form factors of
$\bc\rightarrow\jp$ using analyticity and unitarity constraints on a
particular two-point Green's function and a conformal parameterization
in the manner implemented by Boyd, Grinstein, and Lebed
(BGL)~\cite{Grinstein:2015wqa} for the decays of heavy-light hadrons
to heavy-light or light-light hadrons.  We utilize a slightly
different set of free parameters to simplify the computation for our
particular case of a heavy-heavy meson decaying to another heavy-heavy
meson.  Here we briefly sketch the necessary components, emphasizing
where we differ from the literature.

To derive our constraints, one considers the two-point momentum-space
Green's function $\Pi_J^{\mu \nu}$ of a vectorlike quark current,
$J^\mu \equiv \bar Q \Gamma^\mu Q' \,$.  $\Pi_J^{\mu \nu}$ can be
decomposed in different
ways~\cite{Boyd:1994tt,Boyd:1995cf,Boyd:1995tg,Boyd:1995sq,Boyd:1997kz};
in this work we choose to separate it into spin-1 ($\Pi_J^T$) and
spin-0 ($\Pi_J^L$) pieces {\it \`{a} la}~\cite{Boyd:1997kz} via
\begin{eqnarray}
\Pi_J^{\mu\nu} (q) & \equiv & i \! \int \! d^4 x \, e^{iqx} \left< 0
\left| T J^\mu (x) J^{\dagger \nu} (0) \right| 0 \right> \nonumber \\ 
& = & \frac{1}{q^2} \left( q^\mu q^\nu - q^2
  g^{\mu\nu} \right) \Pi^T_J (q^2) + \frac{q^\mu q^\nu}{q^2} \Pi^L_J
(q^2) \, . 
\label{eq:twopoint}
\end{eqnarray}
From perturbative QCD (pQCD), the functions $\Pi^{L,T}_J$ are known to
contain first- and second-order divergences, respectively, and must
undergo subtractions in order to be rendered finite.  The finite
dispersion relations are:
\begin{eqnarray}\label{eq:chilt}
\chi^L_J (q^2) \equiv \frac{\partial \Pi^L_J}{\partial q^2} & = &
\frac{1}{\pi} \int_0^\infty \! dt \, \frac{{\rm Im} \,
  \Pi^L_J(t)}{(t-q^2)^2} \, , \nonumber \\
\chi^T_J (q^2) \equiv \frac 1 2 \frac{\partial^2 \Pi^T_J}{\partial
  (q^2)^2} & = & \frac{1}{\pi} \int_0^\infty \! dt \, \frac{{\rm Im}
  \, \Pi^T_J(t)}{(t-q^2)^3} \, .
\end{eqnarray}
The freedom to chose a value of $q^2$ can be leveraged to compute
$\chi (q^2)$ reliably in pQCD, far in $q^2$ from where the two-point
function receives nonperturbative contributions from effects such as
bound states and resonances.  The formal condition on $q^2$ to be in
the perturbative regime is
\begin{equation}
(m_Q + m_{Q'})\Lambda_{\rm QCD} \ll (m_Q + m_{Q'})^2 - q^2\,, 
\end{equation}
which, for $Q,Q' = c, b$, $q^2 = 0$ is clearly sufficient.  Existing
calculations of two-loop pQCD $\chi (q^2 \! = \! 0)$ modified by
non-perturbative vacuum contributions~\cite{Generalis:1990id,
Reinders:1980wk,Reinders:1981sy,Reinders:1984sr,Djouadi:1993ss} used
in Ref.~\cite{Boyd:1997kz} can be applied here.  An example of the
state of the art in this regard (although slightly different from the
approach used here) appears in Ref.~\cite{Bigi:2016mdz}.

The spectral functions ${\rm Im} \, \Pi_J$ can be decomposed into a
sum over the complete set of states $X$ that can couple the current
$J^\mu$ to the vacuum:
\begin{equation} \label{eq:FullPi}
{\rm Im} \, \Pi^{T,L}_J (q^2) = \frac 1 2 \sum_X (2\pi)^4 \delta^4
(q - p_X) \left| \left< 0 \left| J \right| \! X \right> \right|^2 \, .
\end{equation}
Each term in the sum is semipositive definite, thereby producing a
strict inequality for each $X$ in Eqs.~(\ref{eq:chilt}).  These
inequalities can be made stronger by including multiple $X$ at once,
as discussed in Refs.~\cite{Boyd:1997kz,Bigi:2017jbd,Jaiswal:2017rve}.
For $X$ we include only below-threshold $\bc$ poles and a single
two-body channel, $\! \bc \! + \! \jp$, implying that our results
provide very conservative bounds.

In contrast to many prior dispersive analyses, $\bc \!
\to \! \jp$, like the $\Lambda_b \! \to \! \Lambda_c$ process studied in
Ref.~\cite{Boyd:1995sq}, does not give the lightest two-body
threshold with the correct quantum numbers; these lighter thresholds
must be taken into consideration.  Depending upon the quantum numbers
indicated by $J$, the first physically prominent two-body
production threshold in $t$ occurs at $B^{(*)} \! + \! D$ (see
Table~\ref{tab:poles}).  In early literature such
as~\cite{Boyd:1995sq}, the branch cut starting at the threshold for
the process of interest was the one used in the dispersive analysis,
while the effect of the cut from the lower threshold up to this
threshold was modeled and argued to amount to a slight loosening of
the unitarity bound given below by Eq.~(\ref{eq:coeffs}).  Here,
however, we represent the analytic features more faithfully by using
the lower threshold directly.  With this fact in mind, we define a
new variable $t_{\rm bd} \! \equiv \!  (M_{B^{(*)}}
\! \!  + \! M_{D})^2$ that corresponds to the first branch point in a
given two-point function, while the $\bc \! + \! \jp$ branch point
occurs at $t_+>t_{\rm bd}$.

With these variables, one maps the complex $t$ plane to the unit disk
in a variable $z$ (with the two sides of the branch cut forming the
unit circle $C$) using the conformal variable transformation
\begin{equation} \label{eq:zdef}
z(t;t_0) \equiv \frac{\sqrt{t_* - t} - \sqrt{t_* - t_0}}
{\sqrt{t_* - t} + \sqrt{t_* - t_0}} \, ,
\end{equation}
where $t_*$ is the branch point around which one deforms the contour,
and $t_0$ is a free parameter used to improve the convergence of
functions at small $z$.  In this mapping, $z$ is real for $t \le t_*$
and a pure phase for $t \ge t_*$.

Prior work that computed the form factors between baryons whose
threshold was above that of the lightest pair in that channel ({\it
i.e.}, $\Lambda_b\rightarrow\Lambda_c$, $\Lambda_b\rightarrow p$) took
$t_*=t_+$~\cite{Boyd:1995tg,Boyd:1997kz}, which introduces into the
region $|z|<1$ a subthreshold branch cut, meaning that the form
factors have complex nonanalyticities that cannot trivially be
removed.  To avoid this issue, we instead set $t_* \! = \! t_{bd}$,
which is possible because we are only interested in the semileptonic
decay region, $m_\ell^2\leq t\leq t_-$, which is always smaller than
$t_{bd}$.  This choice ensures that the only nonanalytic features
within the unit circle $|z| \! = \! 1$ are simple poles corresponding
to single particles $B_c^{(*)+}$, which can be removed by {\it
Blaschke factors\/} described below.  The need to avoid branch cuts
but not poles from $|z| \! < \! 1$ derives from the unique feature of
the Blaschke factors, which can remove each pole given only its
location ({\it i.e.}, mass), independent of its residue.\footnote{The
analytic significance of Blaschke factors for heavy-hadron form
factors was first noted in
Refs.~\cite{Caprini:1994fh,Caprini:1994np}.}  In contrast, correctly
accounting for a branch cut requires knowledge of both the location of
the branch point and the function along the cut.

To remove these subthreshold poles, one multiplies by $z(t;t_s)$
[using the definition of Eq.~(\ref{eq:zdef})], a Blaschke factor,
which eliminates a simple pole $t = t_s$.  Using this formalism, the
bound on each form factor $F_i(t)$ can be written as
\begin{equation} \label{eq:ff}
\frac{1}{\pi} \sum_i\int_{t_{\rm bd}}^\infty \! dt \left|
\frac{dz(t;t_0)}{dt}
\right| \left|P_i(t)  \phi_i (t;t_0) F_i(t) \right|^2 \leq 1 \, .
\end{equation}
The function $P_i(t)$ in Eq.~(\ref{eq:ff}) is a product of Blaschke
factors $z(t;t_p)$ that remove {\em dynamical\/} singularities due to
the presence of subthreshold resonant poles.  Masses corresponding to
the poles that must be removed in $\bc \! \to \! \jp$ are found in
Table~\ref{tab:poles}, organized by the channel to which each one
contributes.  These masses have either been measured by
LHCb~\cite{Aad:2014laa,Aaij:2016qlz} or derived from model
calculations~\cite{Eichten:1994gt}, with uncertainties that are
negligible for our purposes.

\begin{table}
\caption{\label{tab:poles}Lowest $\bc$ states needed for Blaschke
factors with $t \! < \! t_{\rm bc}$ (whose relevant two-body
threshold is indicated by ``Lowest Pair'') for the $J^P$ channels of
interest.  Bold values indicate masses measured by LHCb.}
\begin{center}
\begin{tabular}
{l c c c}
\hline\hline
Type &  $J^P$&Lowest Pair  & $M$ [GeV]\\
\hline
Vector & $1^-$&$BD$&6.337, 6.899, 7.012\\
\hline
Axial & $1^+$ &$B^*D$&6.730, 6.736, 7.135, 7.142\\
\hline
Scalar&$0^+$&$BD$&6.700, 7.108\\
\hline
Pseudoscalar&$0^-$&$B^*D$&\textbf{6.2749(8)}, \textbf{6.842(9)}\\
\hline
\end{tabular}
\end{center}
\end{table}

The weight function $\phi_i(t;t_0)$ is called an {\it outer
function\/} in complex analysis, and is given by
\begin{equation} \label{eq:outer}
\phi_i(t;t_0) = \tilde P_i(t) \left[ \frac{W_i(t)}{|dz(t;t_0)/dt| \,
\chi^j (q^2) (t-q^2)^{n_j}} \right]^{1/2} ,
\end{equation}
where $j \! = \! T,L$ (for which $n_j \! = \! 3,2$, respectively), the
function $\tilde P_i(t)$ is a product of factors $z(t;t_s)$ or
$\sqrt{z(t;t_s)}$ designed to remove {\em kinematical\/} singularities
at points $t = t_s<t_{\rm bc}$ from the other factors in
Eq.~(\ref{eq:ff}), and $W_i(t)$ is computable weight function
depending upon the particular form factor $F_i$.  The outer function
can be reexpressed in a general form for any particular $F_i$ as
 \begin{align}
 \phi_i (t;t_0) = &\sqrt{\frac{n_I}{K \pi \chi}} \,
\left( \frac{t_{\rm bd} - t}{t_{\rm bd} - t_0} \right)^{\frac 1 4} \!
\left( \sqrt{t_{\rm bd} - t} + \sqrt{t_{\rm bd} - t_0} \right)
\left( t_{\rm bc} - t \right)^{\frac a 4} \! \nonumber\\&\times\left(
\sqrt{t_{\rm bd} - t} + \sqrt{t_{\rm bd} - t_-} \right)^{\frac b 2}
\! \left( \sqrt{t_{\rm bd} - t} + \sqrt{t_{\rm bd}} \right)^{-(c+3)}
\! , \label{eq:outer2}
\end{align}
where $n_I$ is an isospin Clebsch-Gordan factor, which is 1 for
$\bc \! \rightarrow \! \jp$.  The remaining factors are found in
Table~\ref{tab:factors}.
\begin{table}
\caption{\label{tab:factors}Inputs entering $\phi_i(t;t_0)$ in
Eq.~(\ref{eq:outer2}) for the meson form factors $F_i$.}
\begin{center}
\begin{tabular}
{l | c c c c c}
\hline\hline
$F_i$ &  $K$ & $\chi$ & $a$ & $b$ & $c$ \\
\hline
$f$   & 24 & $\chi^T(-u)$ & 1 & 1 & 1  \\
${\mathcal F}_1$ & 48 & $\chi^T(-u)$ & 1 & 1 & 2\\
$g$   & 96 & $\chi^T(+u)$ & 3 & 3 & 1  \\
${\mathcal F}_2$ & 64 & $\chi^L(-u)$ & 3 & 3 & 1 \\
\hline
\end{tabular}
\end{center}
\end{table}
Transforming the dispersion-relation inequality based upon
Eq.~(\ref{eq:FullPi}) into $z$-space, Eq.~(\ref{eq:ff}) becomes
\begin{equation} \label{eq:FFrelnz}
\frac{1}{2\pi i} \sum_i\oint_C \frac{dz}{z}
| \phi_i(z) P_i(z) F_i(z) |^2 \le 1 \,
,
\end{equation}
which, upon dividing out the non-analytic terms, allows the expansion
in $z$ corresponding to an analytic function:
\begin{equation} \label{eq:param}
F_i(t) = \frac{1}{|P_i(t)| \phi_i(t;t_0)} \sum_{n=0}^\infty a_{in}
z(t;t_0)^n\, .
\end{equation}
Inserting this form into Eq.~(\ref{eq:FFrelnz}), one finds that the
bound can be compactly written as a constraint on the Taylor series
coefficients:
\begin{equation} \label{eq:coeffs}
\sum_{i;n=0}^\infty a_{in}^2 \leq 1 \, .
\end{equation}
All possible functional dependences of the form factor $F_i(t)$
consistent with Eqs.~(\ref{eq:chilt}) are now incorporated into the
coefficients $a_{in}$.

It is useful to introduce a number of dimensionless parameters that
are functions of the meson masses:
\begin{align}
 r \equiv &\frac{m}{M} , \phantom{xxx}\delta \equiv \frac{m_\ell}{M}
 , \nonumber\\
 \beta \equiv &\frac{M_{B^{(*)}}}{M} , \phantom{xx}\Delta \equiv
\frac{M_{D}}{M} , \nonumber\\
 \kappa \equiv &(\beta+\Delta)^2-(1-r)^2 , \nonumber\\
 \lambda \equiv &(\beta+\Delta)^2-\delta^2 ,
\end{align}
and a parameter $N$ related to $t_0$ in Eq.~(\ref{eq:zdef}) by
\begin{equation} \label{eq:Ndefn}
N \equiv \frac{t_{\rm bd} - t_0}{t_{\rm bd} - t_-} \, .
\end{equation}
It is straightforward to compute the kinematical range for the
semileptonic process given in terms of $z$:
\begin{eqnarray}
  z_{\rm max} & = & \frac{\sqrt{\lambda} - \sqrt{N\kappa}}
  {\sqrt{\lambda} + \sqrt{N\kappa}} \, , \nonumber \\
  z_{\rm min} & = & - \left( \frac{\sqrt{N} - 1} {\sqrt{N} + 1}
  \right) \, , \label{eq:zlimits}
\end{eqnarray}
The minimal (optimized) truncation error is achieved when $z_{\rm min}
= -z_{\rm max}$, which occurs when
\begin{equation} \label{eq:Nopt}
N_{\rm opt} = \sqrt{\frac{\lambda}{\kappa}} \, .
\end{equation}
Evaluating at $N = N_{\rm opt}$, one finds
\begin{equation} \label{eq:zmaxminopt}
  z_{\rm max} = -z_{\rm min} = \frac{\lambda^{1/4} \! -\kappa^{1/4}}
  {\lambda^{1/4} \! +\kappa^{1/4}}\, ,
\end{equation}
From these expressions, we find that the semileptonic decays have
$z_{\rm max,\tau}\approx0.019$ and $z_{\rm max,\mu}\approx0.027$,
where each has a $1.5\%$ variation, depending upon whether the $BD$ or
$B^*D$ threshold is the lowest branch point, $t_{bd}$.

In the limit $t_{\rm bd}\rightarrow t_+$, one obtains
$\Delta\rightarrow r$, $\beta\rightarrow 1$, $\kappa \! \to \! 4r$,
and recovers the expressions in Ref.~\cite{Grinstein:2015wqa}.

\section{Results}\label{sec:results}

Before presenting our bound on $\rjp$, we summarize the constraints
the form factors $g,f,\mathcal{F}_1,\mathcal{F}_2$ are required to
satisfy:
\begin{itemize}
%
%
\item The coefficients $a_n$ of each form factor are constrained by
$\sum_n a_n^2\leq 1$ [Eq.~(\ref{eq:coeffs})], in particular, for the
cases $n \! = \! 1,2$ investigated here.
\item Using Eq.~(\ref{eq:hqss}), the values $g(t_-)$ and
$\mathcal{F}_2(t_-)$ are related to the value of $f(t_-)$, which in
turn is computed from lattice QCD, to within 50\%.
\item All form factors (which are defined to have the same sign
convention as the Isgur-Wise function) are assumed maximal at the
zero-recoil point $t \! = \! t_-$ since the universal form
factor $h$ represents an overlap matrix element between initial and
final states.  Although this condition is not required by the
model-independent parametrization Eq.~(\ref{eq:param}), it appears to
be supported by all the models cited in Table~\ref{tab:models} for
which functional expressions of form factors are provided.  We find
this condition to be suitably implemented via the constraints
$F_i(t_-) \! \geq \!  F_i(0)$ and $\frac{d F_i}{d t}\big|_{t_-} \!
\geq \! 0$, where $F_i$ represents any of the form factors.
\item The relation $\mathcal{F}_1(t_-)=M(1-r)f(t_-)$ [above
Eq.~(\ref{eq:difcof})] is exact.
\item $\mathcal{F}_1(0) \! = \! \frac 1 2 M^2(1-r^2)\mathcal{F}_2(0)$
[above Eq.~(\ref{eq:difcof})] follows from the condition
$A_3(0) \! = \! A_0(0)$.
\end{itemize}
 
Imposing these constraints, we perform our fit in two steps,
reflecting the difference in information between the two form factors
($V, A_1$) for which lattice values have been computed, and the two
($A_0, A_2$) without.
 
\begin{figure}[ht]
\includegraphics[width=\linewidth]{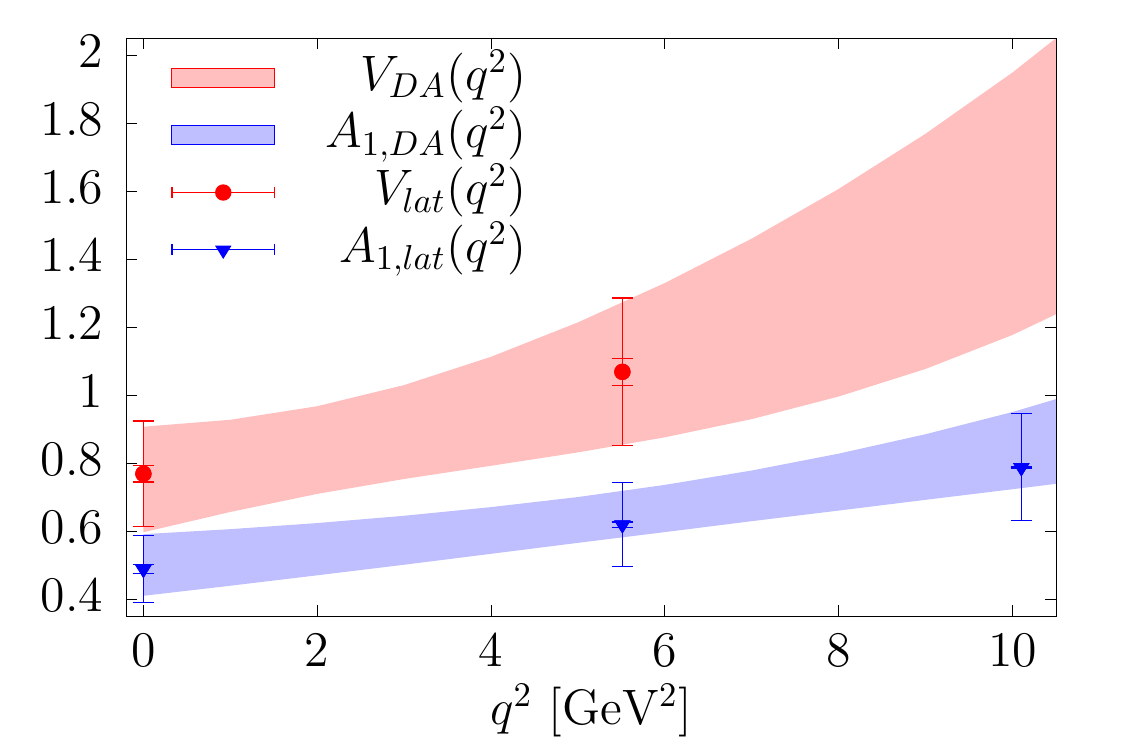}
\caption{\label{fig:latff}$\bc\rightarrow \jp$ form factors $V(q^2)$
(red circles) and $A_1(q^2)$ (blue triangles) from the HPQCD
Collaboration~\cite{Colquhoun:2016osw,ALE}.  The interior bars
represent the statistical uncertainty quoted by HPQCD\@.  The exterior
bars represent the result of including our $f_{\rm lat} \! = \! 20\%$
systematic uncertainty.  The colored bands $DA$ (dispersive analysis)
represent our one-standard-deviation ($1\sigma$) best-fit region.}
\end{figure}

In the first step, random Gaussian-distributed points are sampled for
the form factors $g$ and $f$ [equivalently, by Eq.~(\ref{eq:FFrelns}),
$V$ and $A_1$] whose mean gives the HPQCD results.  The combined
uncertainties are given by the quadrature sum of the reported
uncertainty $\delta_{\rm lat}$ of the form-factor points and an
additional systematic uncertainty, $f_{\rm lat}$ (expressed as a
percentage of the form-factor point value) that we use to estimate the
uncomputed lattice uncertainties ({\it i.e.}, finite-volume
corrections, quark-mass dependence, discretization errors).  $f_{\rm
lat}$ is taken to be 1, 5, or 20\% of the value of the form factor
from the lattice.  For our final result, we suggest using $f_{\rm lat}
\! = \! 20\%$, while the other two values are helpful for
understanding future prospects with improved lattice data.  Using
these sample points, we compute lines of best fit, from which we
produce the coefficients $a_n$.  The resulting bands of allowed form
factors are shown for $f_{\rm lat} \! = \! 20\%$ in
Fig.~\ref{fig:latff}, alongside the HPQCD results.

\begin{figure}[ht]
 \includegraphics[width=\linewidth]{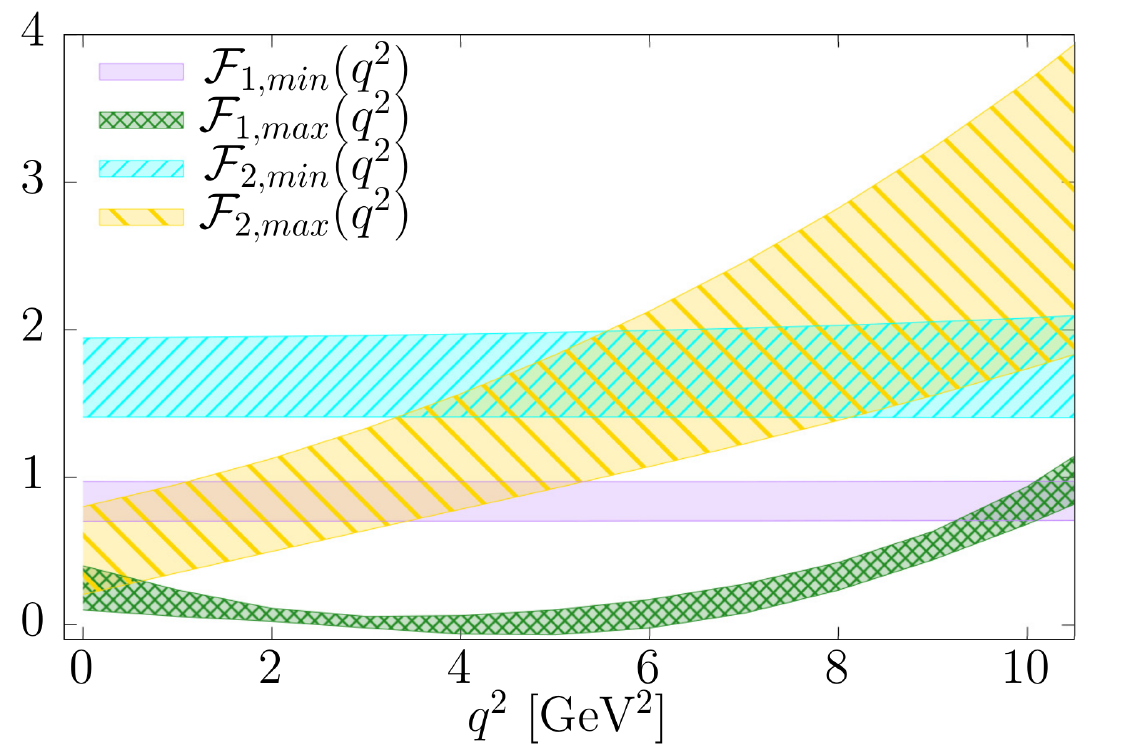}
 \caption{\label{fig:fsff}Dimensionless form factors $\mathcal{F}_1/[
 \frac 1 2 M^2 (1 - r^2)]$ and $\mathcal{F}_2$ that provide the
 maximum and minimum $\rjp$ values consistent with lattice and
 heavy-quark spin-symmetry constraints.  The colored bands represent
 the $1\sigma$ range due to the uncertainty associated with the HPQCD
 results, combined with an $f_{\rm lat} \! = \! 20\%$ systematic
 uncertainty.}
\end{figure}

In the second step, we compute $\mathcal{F}_1$ and $\mathcal{F}_2$
(which include $A_{0,2}$), for which no lattice information exists.
One could adopt the tactic of randomly generating points with some
prior distribution, which, once accounting for the constraints, could
be used to suggest a mean value of $\rjp$ with some prior-dependent
uncertainty.  We instead opt to remove this possible dependence by
obtaining the numerical maximum and minimum $\rjp$ values, subject to
the computed $f,g$ values and the constraints listed above.  In this
way, the only uncertainties included are those from the lattice-QCD
results and the violations of the heavy-quark spin-symmetry relations.
The resulting bands of form factors for $\mathcal{F}_1$ and
$\mathcal{F}_2$ that produce the minimum and maximum values of $\rjp$
subject to the constraints are plotted in Fig.~\ref{fig:fsff}.

Having computed all the form factors, we present the 95\% CL ranges
for $\rjp$ as a function of the truncation power $n=1,2$ in the
dispersive analysis coefficients of Eq.~(\ref{eq:param}) and the
$1,5,20\%$ systematic uncertainty $f_{\rm lat}$ associated with the
lattice data.  The full results are presented in
Table~\ref{tab:rvalues}\@.  The bound on $\rjp$ appears relatively
insensitive to increasing the number of free parameters $a_n$, and
only mildly dependent upon $f_{\rm lat}$.  One might be concerned that
increasing $n$ could dramatically change these results, but we note
that the typical value of $\sum_n a_n^2$ for $n \! = \! 1$ is ${\cal
O}(10^{-2})$, while for $n \! = \! 2$ we find $a_2\approx 1$.  While
the dispersive constraint is saturated in the $n \! = \! 2$ case, the
bound on $\rjp$ is only enlarged by $5\%$.  However, the
saturation of any particular $a_n$ is not necessary to find the effect
of higher $a_n$ to be negligible.  Since higher-order terms are
suppressed by $z_{\rm max}\approx 0.03$, in order for these terms to
contribute strongly, one must have $a_{n+1} z_{\rm max} \! \gtrsim \!
a_n $.  Such an $a_{n+1}$ value would either have to
violate saturation $\sum_n a_n^2\leq 1$ once the lower-order terms
$a_n$ are fixed, or else it would change the numerical
results very little.

\begin{table}
\caption{\label{tab:rvalues}95\% CL upper and lower bounds on
$R_{\jp}$ as a function of the truncation power $n$ of coefficients
included from Eq.~(\ref{eq:param}) and the systematic lattice
uncertainty $f_{\rm lat}$.}
\begin{center}
\begin{tabular}
{c c c}
\hline\hline
$f_{\rm lat}$ & $n=1$ \ & $n=2$\\
\hline
1 & [0.21, 0.33] & [0.20, 0.35]\\
5 & [0.20, 0.33] & [0.20, 0.35]\\
20 & [0.20, 0.36] & [0.20, 0.39]\\
\hline
\end{tabular}
\end{center}
\end{table}

\begin{figure}[ht]
\includegraphics[width=\linewidth]{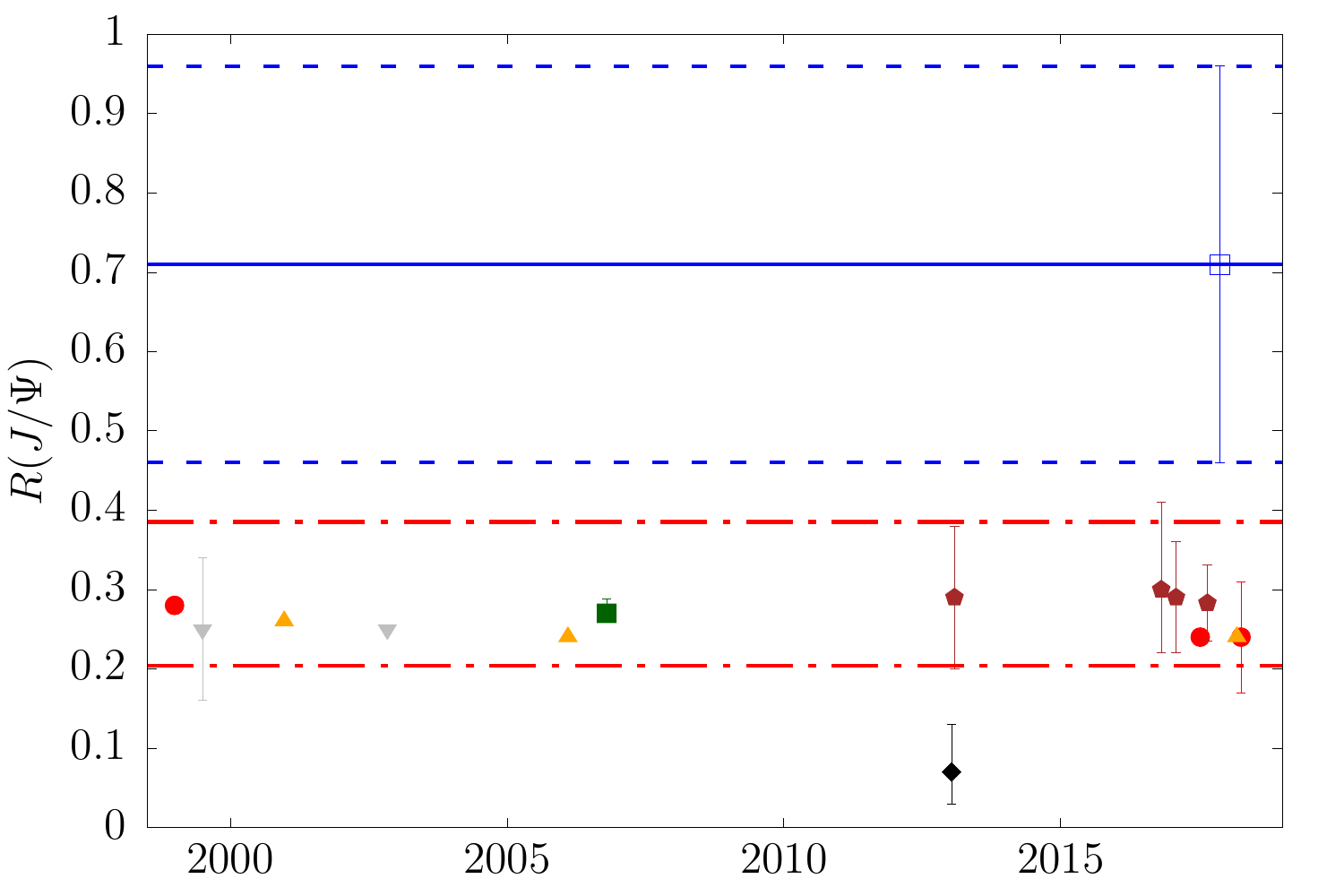}
\caption{\label{fig:rjp}$R(\jp)$ from the LHCb experiment (blue
open square, $1\sigma$ uncertainty denoted by blue dashed lines), our
bound (red dash-dotted lines), and models (points colored by model
type), as listed in Table~\ref{tab:models}.}
\end{figure}

In Fig.~\ref{fig:rjp} we plot the previous model-dependent values of
$\rjp$ alongside the LHCb result and our 95\% CL bound of
$0.20\leq\rjp\leq0.39$, as a function of publication date.  One can
see that, while many of the previous model results lie within our 95\%
CL band, some are either partially or entirely excluded.  The
anomalously low NRQCD result of Ref.~\cite{Qiao:2012vt} is in severe
disagreement with our bound (the small $\rjp$ of
Ref.~\cite{Qiao:2012vt} can be attributed to a larger-than-typical
muonic branching ratio), while all the other models that have
included $1\sigma$ theory uncertainty estimates
are seen to remain compatible with the dispersive bounds
obtained from a fairly sparse set of lattice results.  Our 95\% CL
band should be viewed as a model-independent upper limit (subject to
the assumptions listed above) for the largest theory uncertainty any
model can allow and remain consistent with analyticity, unitarity, and
existing lattice ``data.''  With better lattice results---and/or
actual experimental measurements of the form factors at any values of
$q^2$---the allowed parameter space for any given model will become
severely curtailed.

Considering that the fits to the lattice data already constrain the
form factors $g$ and $f$ well, we can ask what new piece of
information would most improve our bounds.  Since $\mathcal{F}_2$
essentially affects only the $\tau$ channel, it is obvious that this
form factor is the one most important to reducing the range in $\rjp$.
By inspection of both $\mathcal{F}_1$ and $\mathcal{F}_2$ in
Fig.~\ref{fig:fsff}, we find that the $q^2$ dependence of the form
factors generating the maximum and minimum $\rjp$ values are quite
different in shape.  The minimum-$\rjp$ ones clearly prefer flat form
factors, while the maximum ones are nearly zero at $q^2 \! = \! 0$ and
rise to their maximum allowed values at $q^2 \! = \! t_-$.  This
dependence suggests that obtaining a value of $\mathcal{F}_2(q^2 \! =
\! 0)$ would greatly improve the lower bound, while
$\mathcal{F}_2(t_-)$ would restrict the upper bound.  With the LHCb
result already lying slightly above our bound, it would be most
incisive to reduce our upper bound.  This zero-recoil form factor is
directly related by $\mathcal{F}_2(t_-)=2A_0(t_-)$ to a traditional
lattice form factor, and therefore should be possible to compute.
 
To investigate the possible effect of this new information, we
consider a synthetic point $\mathcal{F}_2(t_-) \! = \! 2 (1 \! \pm \!
f_{\rm lat})$.  This particular value is chosen because it lies near
the average of the minimum and maximum preferred values, and is
similar to the values suggested in models.  Taking $f_{\rm lat} \! =
\! 20\%$, we find that the bound could tightened to [0.20,\,0.35].  If
this point and the existing 5 lattice points reached $f_{\rm lat} \! =
\! 1\%$, one could anticipate a range of [0.21,\,0.32].  So, an
additional lattice point at $\mathcal{F}_2(t_-)$ could improve the
bound by the same amount as reducing the uncertainty $f_{\rm lat}$
from 20\% to 1\% (as seen in Table~\ref{tab:rvalues}), but with
substantially less computing resources.

\section{Discussion and Conclusion}\label{sec:con}

In contrast to the model-dependent previous works, we have
presented a model-independent bound on $\rjp$, finding it constrained
to lie in the range $0.20\leq\rjp\leq0.39$ at the 95\% CL\@.  At this
level, we find that the LHCb result is consistent with the Standard
Model at $1.3 \, \sigma$.  The near-term outlook for a
higher-statistics LHCb measurement, coupled with new lattice results,
promises to reduce the uncertainty on the experimental and theoretical
values dramatically.

Even without a lattice QCD calculation of the $\mathcal{F}_2$ form
factor, additional potential areas of improvement can be investigated.
Experience in the heavy-light sector and the fact that the $\rjp$
bounds require saturating $\sum a_n^2=1$ suggest that including
multiple states appearing in the dispersion relation can provide
complementary information to help constrain the form factors further,
and in this case one can additionally include the lattice results for
$B\rightarrow D^{(*)}$~\cite{Lattice:2015rga,Na:2015kha,
Bailey:2014tva,Harrison:2016gup,Harrison:2017fmw,Bailey:2017xjk} and
$\Lambda_b\rightarrow \Lambda_c$~\cite{Detmold:2015aaa}.

\begin{acknowledgments}
This work was supported by the U.S.\ Department of Energy under
Contract No.\ DE-FG02-93ER-40762 (T.D.C.\ and H.L.) and the National
Science Foundation under Grant No.\ PHY-1403891 (R.F.L.).
\end{acknowledgments}
 
\bibliographystyle{JHEP}
\bibliography{wise}
\end{document}